\documentstyle[11pt,newpasp,twoside,epsf]{article}
\def\edcomment#1{\iffalse\marginpar{\raggedright\sl#1\/}\else\relax\fi}
\reversemarginpar

\begin{document}
\title{The orbital evolution of planets in disks}
 \author{Wilhelm Kley}
\affil{Max-Planck-Institut f\"ur Astronomie,
K\"onigstuhl 17, D-69117 Heidelberg, Germany}

\begin{abstract}
The orbital parameters of the observed extrasolar planets differ
strongly from those of our own solar system. The differences
include planets with
high masses, small semi-major axis and large eccentricities.
We performed numerical computations of embedded planets in disks
and follow their mass growth and orbital evolution over several
thousand periods.

We find that planets do migrate inwards on timescales of about $10^5$ years
on nearly circular orbits, during which they may grow up to about
5 Jupiter masses.
The interaction of the disk with several planets may halt the migration
process and lead to a system similar to the solar planetary system.
\end{abstract}
\section{The model}
The evolution of protoplanets can only be studied simultaneously with
that of the protostellar disk in which the planets are still embedded
early in their lifetime.
The disk is assumed to be flat
and non-self gravitating, and is modelled by the planar two-dimensional
($r-\varphi$) Navier-Stokes equations.
The mutual gravitational interaction of the planets and the star,
and the gravitational torques of the disk acting
on the planets and the central star are included.
The time dependent hydrodynamic equations are integrated by using
a finite difference scheme (Kley 1998, 1999). The computations cover
the radial range from 1.2 to 20~AU.
The initial surface density distribution is axisymmetric and
follows $\Sigma \propto r^{-1/2}$; to speed up the computations
an annular gap is imposed (Kley 1999). The material orbits the star 
with Keplerian speed.
The planets of one Jupiter mass
are initially placed at typical distances of several AU from
the star. The whole system (disk, star and planets) is then evolved
for several thousand orbital periods.
\section{Results of a single Planet}
The presence of the planet perturbs the axisymmetric
density by generating trailing spirals (Fig.~1).
In the situation where the tidal torques are greater than the internal viscous
torques in the disc, and the disc response becomes non linear,
an annular gap, or surface density depression, will form at the radius
of the planet. The radial width of the gap is determined by the
equilibrium of gravitational (gap openeing) torques and viscous (gap
closing) torques. For a one Jupiter mass planet and a typical disk viscosity
of $\alpha \approx 10^{-3}$ the width of the gap is about two
Roche-lobe sizes of the planet.

Although the density around the planet is greatly reduced in the gap region,
material from upstream
may enter the Roche-lobe and eventually be accreted onto the planet (Fig.~2).
Typical values for the viscosity ($\alpha \approx 10^{-3}$)
and mass of the disk ($M_{disk}=0.01 M_\odot$) yield a mass
accretion rate onto the planet of
$4 \times 10^{-5} M_{Jup} /yr$ (Kley 1999).
\begin{figure}[h]
\plotone{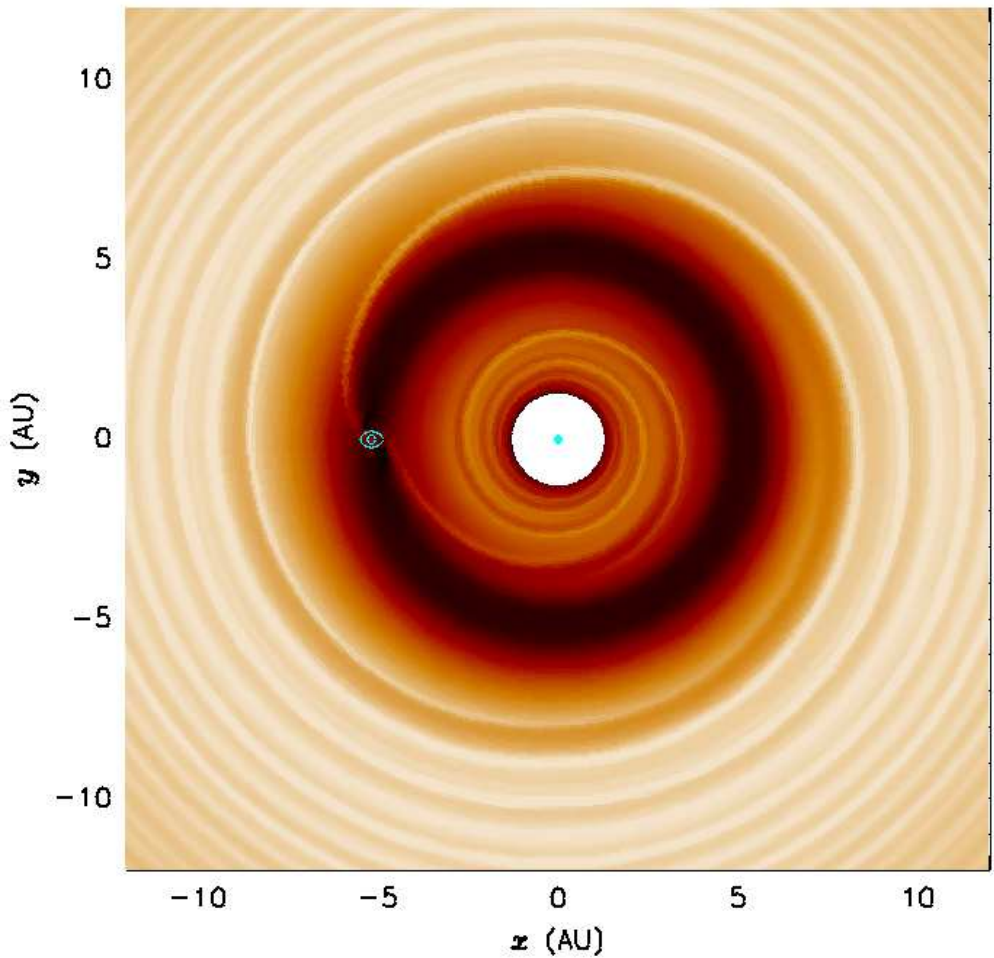}
\caption{
Gray scale plot of the surface density
after 200 orbits of the planet, which is located at $x=-5.2, y=0.0$.
The gravitational torques due to the planet acting on the disk
lead to the excitation of spiral density waves.
}
\end{figure}
The rate increases with larger disk viscosity and decreases
when the mass of the planet grows.
For very low disk viscosity and larger planetary masses
the mass accumulation eventually terminates, and the maximum mass the planet
may reach is about 5 Jupiter masses ($M_{Jup}$),
see also Bryden et al. (1999), and Nelson et al. (2000).

The material in the disk acts gravitationally on the
planet and disturbs the orbit.
The net torque onto the planet,
obtained by summation over the different annuli, is
negative which leads to an
inward migration on typical timescales of $10^5 yrs$ (Nelson et al. 2000).

\begin{figure}
\plotone{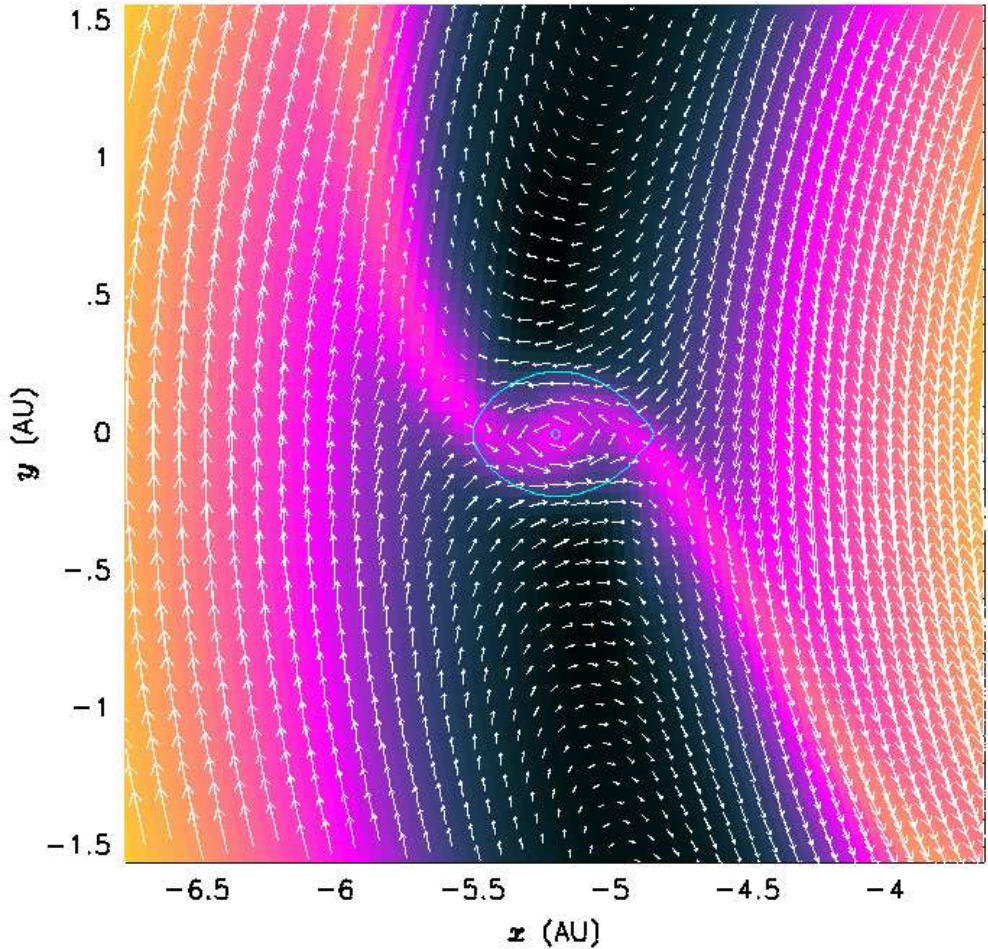}
\caption{
The flow field in the vicinity of the embedded
protoplanet in the corotating frame. The blue line indicates the
size of the Roche-lobe of the planet.
}
\end{figure}
\section{Evolution of two planets}
As it is believed that more eccentric orbits are caused by the interaction
of several objects, we have performed a run with two embedded Jupiter-type
planets. Initially they were located at $1 a_J$ and $2 a_J$ in opposition
to each other, i.e. separated by $\Delta \varphi = 180^{\circ}$.
The two planets were assumed to be on circular orbits initially.

Two planets create a much more complicated
pattern of wave-like disturbances
in the density distribution of the disk than just one planet does.
In a calculation of one planet on a fixed circular orbit,
the wave pattern induced in the disk is stationary
in the co-rotating frame. As seen in Fig.~3, in case of two planets
(the Roche-lobes are indicated) it changes strongly with time.
As both planets accrete essentially all of the mass which enters
their Roche-lobes, their masses grow in time.
The material located initially radially between the planets is
consumed by the planets within the first few hundred orbital periods.

\begin{figure}[h]
\plotone{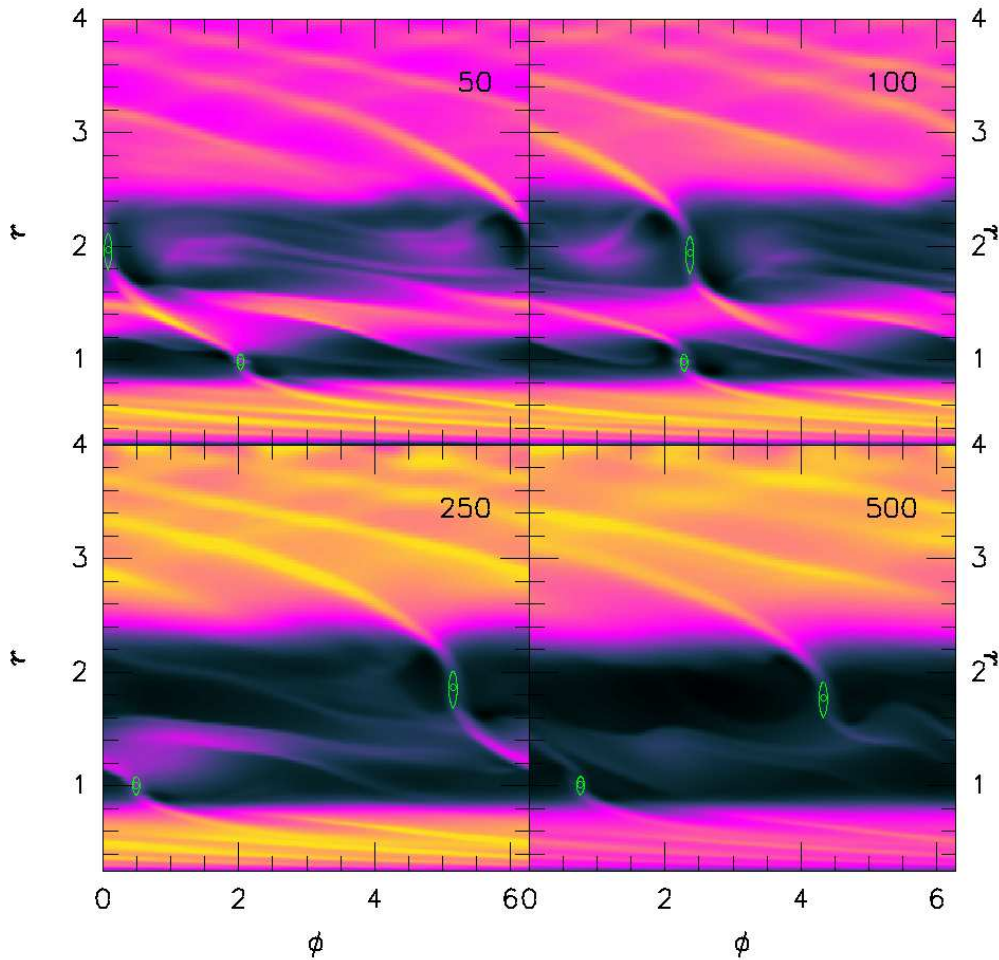}
\caption{
Surface density distribution in $r-\varphi$ display at four 
evolutionary times.
The panels are labelled consecutively by
the number of orbits of the inner planet.
}
\end{figure}
Finally, for the inner planet, the only gas available is that
which flows through the gap created by the outer planet.
Hence the mass of the inner planet grows at a smaller rate than the outer
one, see Fig.~4 (bottom).
At the same time, the gravitational interaction of the two planets, the star
and the disk lead to changes in the orbital elements of the objects.
The change in the semi-major axis of each planet is given in
Fig.~4 (top).
During the evolution, the fluctuations of the radial distances to the
star strongly increase, indicating a growth in eccentricity of the planet.
\begin{figure}[h]
\plotone{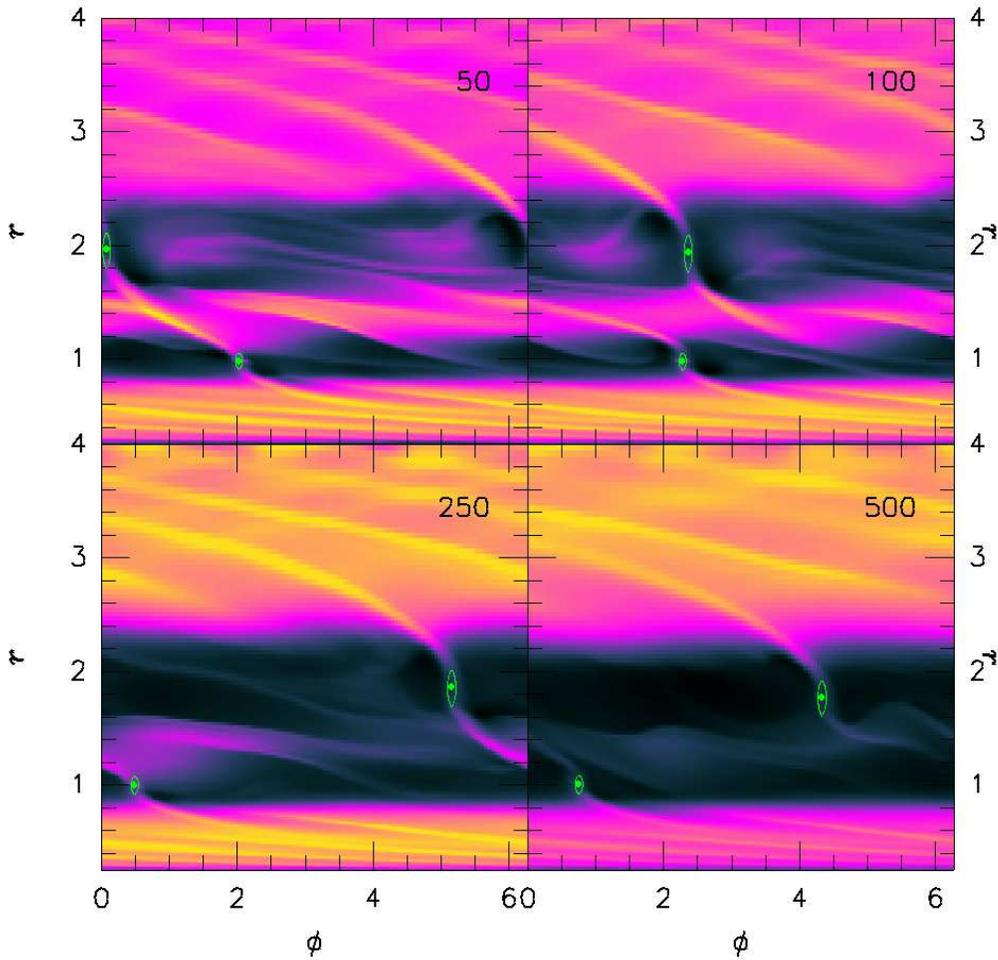}
\caption{
{\bf Top:}
Evolution of the semi-major axis of the two planets.
{\bf Bottom:}
Evolution of the masses of the two planets
(1: inner planet, 2: outer planet).
The inferred mass accretion rates
during the long term evolution are indicated.
Time is given in units of the initial period of the inner planet.
}
\end{figure}  
\section{Conclusions}
By initially considering one planet on a fixed circular orbit, it was shown
that even though the planet opens up an annular gap in disk it is
nevertheless able to accrete more mass from its surroundings.
As more massive planets induce a wider and deeper gap the mass
accumulation essentially terminates,
which puts the upper limit to the mass of the planet mass at about
5 to 10 $M_{Jup}$, in good agreement with the observations.

By considering the evolution of a system consisting of
two Jupiter-type planets, we show (Kley 2000) that by mutual gravitational
interaction the inward motion of the inner planet may come to a halt.
As the outer planet continues to move inward and approaches the inner,
the gravitational interaction between the two planets increases.
The resulting configuration may be unstable, and lead to systems
similar to $\upsilon$~And.

If the protoplanetary nebula has dissipated before
the planets come very close to each other, one is left with a system of
massive planets at several AU distance, similar to our own Solar System.

\end{document}